\documentclass[aps,prb,twocolumn,superscriptaddress]{revtex4}
\usepackage{amsmath}    
\usepackage{graphicx}   

\def\U#1{{%
\def\O{\mbox{O}}
\def\u{\mbox{u}}
\mathcode`\u=\mu
\mathcode`\O=\Omega
\mathrm{#1}}}
\def\ii{{\mathrm{i}}}
\def\ee{{\mathrm{e}}}

\def\bra#1{\langle #1|}
\def\ket#1{|\mbox{$#1$}\rangle}

\def\bracketi#1#2{\langle\mbox{$#1$}|\mbox{$#2$}\rangle}
\def\bracketii#1#2#3{\langle\mbox{$#1$}|\mbox{$#2$}|\mbox{$#3$}\rangle}
\begin{document}

\title{Observation of Geometric Phases in Quantum Erasers}

\author{H. Kobayashi}
\affiliation{Department of Electronic Science and Engineering, Kyoto
University, Kyoto 615-8510, Japan}
\author{S. Tamate}
\affiliation{Department of Electronic Science and Engineering, Kyoto
University, Kyoto 615-8510, Japan}
\author{T. Nakanishi}
\affiliation{Department of Electronic Science and Engineering, Kyoto
University, Kyoto 615-8510, Japan}
\author{K. Sugiyama}
\affiliation{Department of Electronic Science and Engineering, Kyoto
University, Kyoto 615-8510, Japan}
\author{M. Kitano}
\affiliation{Department of Electronic Science and Engineering, Kyoto
University, Kyoto 615-8510, Japan}

\begin{abstract}
In this study, we report the manifestation of geometric phases in the setup for
 quantum erasers. 
Our experiment includes a double-slit
 interferometer with the polarization as an internal
 state of a photon. 
With regard to the visibility of the interference fringe, we can
 demonstrate the disappearance of fringes by which-path marking and
 the recovery of interference using quantum erasers, and the phase shift
 of the fringe due to the evolution of the polarization state is attributed to
 the geometric phase or the Pancharatnam phase.
For a certain arrangement, the geometric phase can be very sensitive to
 a change in state and this is observed as a rapid displacement of the
 fringes.
\end{abstract}

\maketitle

\section{Introduction}

Wave-particle duality
is one of the most
intriguing features of quantum mechanics. 
This property manifests itself prominently in the Young's double slit
experiment; each quanta creates a single spot on the observation
plane according to the probability amplitude, 
and the spots created by
thousands of quanta result in a clear fringe pattern 
due to the superposition of wavefunctions for the two possible paths followed 
by the quantum particle
\cite{parker72:_singl_photon_doubl_slit_inter,rueckner96:_lectur_demon_of_singl_photon_inter}. 
Here, we assume that there exists a device to ``mark'' 
each particle according
to the path followed by it. This operation, called as which-path
marking, enables us to distinguish the two states for the path so that 
the superposition of the path states has been collapsed and the interference
disappears.

Surprisingly, although the which-path marking destroys the interference,
we can recover the interference fringe by erasing 
the which-path information.
This idea called as the quantum eraser was first proposed by
Scully and Dr\"{u}hl\cite{scully82:_quant_eraser}, and it has been
 discussed extensively in connection with the wave-particle duality
\cite{tan93:_loss_of_coher_in_inter,
herzog95:_compl_and_quant_eraser,engert96:_fring_visib_and_which_way_infor,
luis98:_compl_enfor_by_random_class_phase_kicks,bjork98:_compl_and_quant_erasur_in,
schwindt99:_quant_wave_partic_dualit_and,walborn02:_doubl_slit_quant_eraser}.

A simple demonstration of the path marking and the quantum eraser using
the internal states of a photon can be demonstrated using a double-slit
interferometer as follows. 
A photon is marked with the right and left circular polarization states according
to the paths.
Because we could distinguish the path state by measuring the polarity of the
circular polarization, no interference pattern is observed.
However, when a linear polarizer is placed behind the double slit, the
circular polarizations are projected into the same linear polarization
and which-path information in the polarization state is completely erased. Therefore, the
interference fringes are recovered\cite{walborn03:_quant_erasur,hillmer07:_do_it_yousel_quant_eraser}. 

In addition to the recovery of interference, due to the change in the
polarization states, the quantum eraser also induces 
an additional phase shift determined by 
three polarization states: two states
due to the which-path marking and one due to the linear
polarizer used for the quantum eraser. 
This phase shift is called as the Pancharatnam
phase, which is proportional to the area of the spherical
triangle connecting the three states on the Poincar\'{e}
sphere\cite{pancharatnam56:_proc,aravind92:_simpl_proof_of_panch_theor}. 
From this geometric property, it is also called as the geometric phase
\cite{berry84:_quant_phase_factor_accom_adiab_chang,berry87:_adiab_phase_and_panch_phase}.
It has been shown that the Pancharatnam phase can be very sensitive to
a change in state for a certain arrangement
\cite{schmitzer93:_nonlin_of_panch_topol_phase,tewari95:_four_arm_sagnac_inter_switc,bhandari97:_polar_of_light_and_topol_phases,hils99:_nonlin_of_pnach_geomet_phase,li99:_exper_obser_of_nonlin_of}.

Recently, based on the interferometric point of view, Tamate and his
co-workers revealed that the Pancharatnam phase contributes to the weak
measurements\cite{aharonov88:_how_resul_of_measur_of,tamate09:_geomet_aspec_of_weak_measur}. 
In weak measurements, we can obtain unusual results that lie well outside of the range of eigenvalues of an
observable. 
Owing to this property, the weak measurement is very useful for experimentally detecting 
minute effects\cite{hosten08:_obser_of_spin_hall_effec,dixon09:_ultras_beam_deflec_measur_via}. 
It has been shown that the high sensitivity of the Pancharatnam phase to a change in state plays an
essential role in weak measurements\cite{tamate09:_geomet_aspec_of_weak_measur}.

\begin{figure}[tbp]
\begin{center}
\includegraphics[width=8cm]{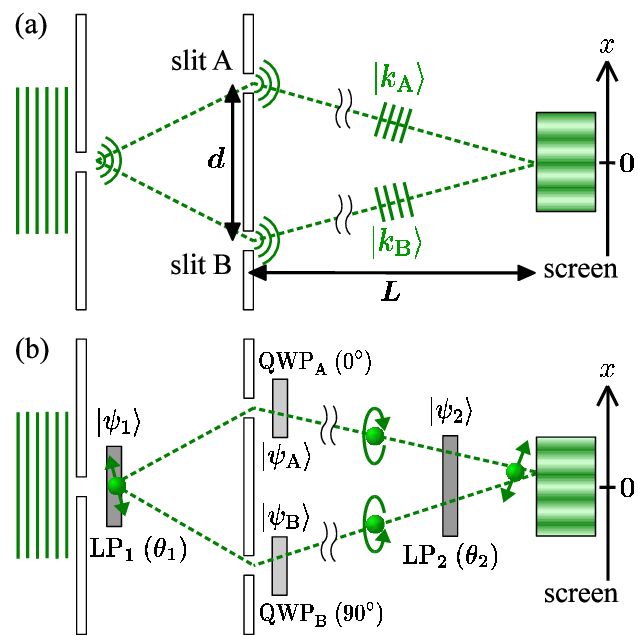}
\caption{(Color online). (a) A typical Young's double-slit interferometer using photons. 
The path states of the photon are represented as the
 state of the transverse wave numbers, $\ket{k_\U{A}}$ and
 $\ket{k_\U{B}}$. (b) A double-slit interferometer with an internal
 state. The which-path
 marker comprises of the linear polarizer LP$_1$ and the two
 quarter-wave plates QWP$_\U{A}$ and QWP$_\U{B}$. The linear polarizer
 LP$_2$ serves as the quantum eraser. The states $\ket{\psi_1}$, $\ket{\psi_2}$,
 $\ket{\psi_\U{A}}$, and $\ket{\psi_\U{B}}$ are the polarization states
 after LP$_1$, LP$_2$, QWP$_\U{A}$, and QWP$_\U{B}$, respectively. 
The angles of the transmission axes of LP$_1$ and LP$_2$ are $\theta_1$ and
 $\theta_2$, and those of the fast axes of QWP$_\U{A}$ and QWP$_\U{B}$ are
 $0^\circ$ and $90^\circ$, respectively.}
\label{fig:double-slit}
\end{center}
\end{figure}
In this study, we report the manifestation of the Pancharatnam phase in the
setup for quantum erasers. 
The loss of interference by
which-path marking can be explained by the fact that 
the interference pattern destroyed by the which-path marking contains
two complete interference patterns that are shifted by different
amounts due to the Pancharatnam phase. 
This is demonstrated in our experiment for a double-slit
interferometer with internal states of a photon. 
With regard to the visibility of the interference fringe, we can
confirm the disappearance of interference due to which-path marking and
recovery of interference using the quantum eraser. 
On the other hand, we can observe the nonlinear variation of 
the Pancharatnam phase 
with regard to the phase shift of the fringe in the same setup. 
Although each phenomenon
has already been described in previous works, 
in our experiment we can observe these phenomena using a single setup.
It may be useful for showing a unified viewpoint of the quantum
eraser and the geometric phase, and based on this viewpoint, our experiment can be 
interpreted as the minimal setup of weak measurement for 
the measured qubit (polarization states) coupled with the qubit meter
(path states)\cite{brun08:_test_of_weak_measur_two,wu09:_weak_measur_with_qubit_meter}.

The remainder of this paper is organized as follows.
In Sec.~\ref{sec:double-slit-quantum}, we introduce a theoretical model
for a double-slit interferometer with the internal states and analyze
the interference pattern in the process of which-path marking and the
quantum eraser. Moreover, we confirm the nonlinear variation of the Pancharatnam phase in a certain arrangement.
In Sec.~\ref{sec:experiments}, we describe our experimental setup and results on
the quantum eraser and the Pancharatnam phase. 
A summary is presented in Sec.~\ref{sec:summary-discussion}.

\section{Theoretical analysis of double-slit interferometers with internal states}
\label{sec:double-slit-quantum}
An interferometer with internal states can be analyzed as a quantum system composed of the path state and the
internal state. 
In this section, 
we theoretically analyze the interference patterns in our
double-slit experiment with regard to both the intensity and the phase.

\subsection{Which-path marking}
Because of the large distance between the double slit and the screen,
the state of the photon through slits $\U{A}$ and $\U{B}$ can be assumed to be an eigenstate of
the transverse wave
numbers on the screen, $\ket{k_\U{A}}$ and $\ket{k_\U{B}}$, respectively,
as shown in Fig.~\ref{fig:double-slit}(a). These states
satisfy the normalization condition 
$\bracketi{k}{k^\prime}=\delta(k-k^\prime)$, where $\delta(\cdot)$ shows
the Dirac delta function.
In our setup, a photon is marked with the polarization states $\ket{\psi_\U{A}}$ and
$\ket{\psi_\U{B}}$ according to the paths using the quarter-wave plates
QWP$_\U{A}$ and QWP$_\U{B}$ (see Fig.~\ref{fig:double-slit}(b)). 
Assuming that the photon has a 50:50 chance of passing through each slit, 
the total state vector for the composite system can be represented as
the following superposition:
\begin{align}
\ket{\Psi_\U{m}}=\ket{\psi_\U{A}}\ket{k_\U{A}}+\ket{\psi_\U{B}}\ket{k_\U{B}}.
\label{eq:16}
\end{align}
Here, the path states and two polarization states are correlated or
\textit{entangled}.
We introduce the operator $\hat{P}_x\equiv\hat{I}\otimes\ket{x}\bra{x}$, 
which projects the path state into the position state $\ket{x}$ on the screen.
With the position representation of the wave-number
eigenfunction, $\bracketi{x}{k}=\ee^{\ii kx}/\sqrt{2\pi}$, 
the probability distribution $P_\U{m}(x)$ is given by
\begin{align}
P_\U{m}(x)&=\bracketii{\Psi_\U{m}}{\hat{P}_x}{\Psi_\U{m}}\nonumber\\
&\propto 1+V_\U{m}
\cos\left(kx-\delta_\U{m}\right),
\label{eq:23}
\end{align}
where $k\equiv k_\U{B}-k_\U{A}$ and
\begin{align}
V_\U{m}&=\left|\bracketi{\psi_\U{B}}{\psi_\U{A}}\right|,\label{eq:9}\\
\delta_\U{m}&=\arg\bracketi{\psi_\U{B}}{\psi_\U{A}}.
\label{eq:8}
\end{align}
For the double-slit apparatus, $k$ is calculated as $k=2\pi d/\lambda L$,
where $\lambda$ is the wavelength of light; $d$, the distance between
two slits; and $L$, the distance between the double slit and the screen.
The coefficient of the interference term, $V_\U{m}$, can be
experimentally obtained from the fringe pattern as the visibility
\begin{align}
V_\U{m}=\frac{P_\U{max}-P_\U{min}}{P_\U{max}+P_\U{min}},
\label{eq:15}
\end{align}
where $P_\U{max}$ and $P_\U{min}$ are the maximum and minimum values
of $P_\U{m}(x)$, respectively.

The degradation of visibility is related to the efficacy of which-path
marking, which depends on the inner product
$\bracketi{\psi_\U{A}}{\psi_\U{B}}$. 
The lesser the value of $|\bracketi{\psi_\U{A}}{\psi_\U{B}}|$, the
lesser is the visibility. 
In particular, when $\bracketi{\psi_\U{A}}{\psi_\U{B}}=0$, 
two states are perfectly distinguishable and the path
followed by the photon is discriminated unambiguously. 
Then, the interference is completely eliminated.

\subsection{Quantum eraser}
Now, we erase the which-path information by the projection 
of polarization using the linear polarizer LP$_2$ that projects
the polarization state into $\ket{\psi_2}$ (see
Fig.~\ref{fig:double-slit}(b)). 
The state vector for the composite system after LP$_2$ is calculated 
as follows:
\begin{align}
\ket{\Psi_\U{f}}&=\ket{\psi_2}\bracketi{\psi_2}{\Psi_\U{m}}
=\ket{\psi_2}
\Bigl(c_\U{A}\ket{k_\U{A}}
+c_\U{B}\ket{k_\U{B}}\Bigr),
\label{eq:11}
\end{align}
where $c_\U{A}=\bracketi{\psi_2}{\psi_\U{A}}$ and 
$c_\U{B}=\bracketi{\psi_2}{\psi_\U{B}}$.
The probability distribution $P_\U{f}(x)$ is
given by
\begin{align}
P_\U{f}(x)&=\bracketii{\Psi_\U{f}}{\hat{P}_x}{\Psi_\U{f}}\nonumber\\
&\propto
1+V_\U{f}\cos\left(kx-\delta_\U{f}\right),
\label{eq:24}
\end{align}
where the visibility $V_\U{f}$ and the phase shift $\delta_\U{f}$ are
given as
\begin{align}
V_\U{f}&=\frac{2|c_\U{A}|\cdot|c_\U{B}|}
{|c_\U{A}|^2+|c_\U{B}|^2},
\label{eq:25}\\
\delta_\U{f}&=
\arg\bracketi{\psi_\U{B}}{\psi_2}\bracketi{\psi_2}{\psi_\U{A}}.\label{eq:10}
\end{align}
Equation~(\ref{eq:25}) shows that even when $\ket{\psi_\U{A}}$ is orthogonal to
$\ket{\psi_\U{B}}$, the visibility is recovered completely provided that
$|c_\U{A}|=|c_\U{B}|$.
In this case, the states of the which-path marker, $\ket{\psi_\U{A}}$ and
$\ket{\psi_\U{B}}$, are projected into the same polarization state
$\ket{\psi_2}$ with the same probability, and it cannot be
determined whether the photon came
from slit A or B. 
This implies that LP$_2$ completely erases the which-path information, and
the interference is recovered. 

\subsection{Quantum eraser and Pancharatnam phase}
\begin{figure}
\begin{center}
\includegraphics[width=7cm]{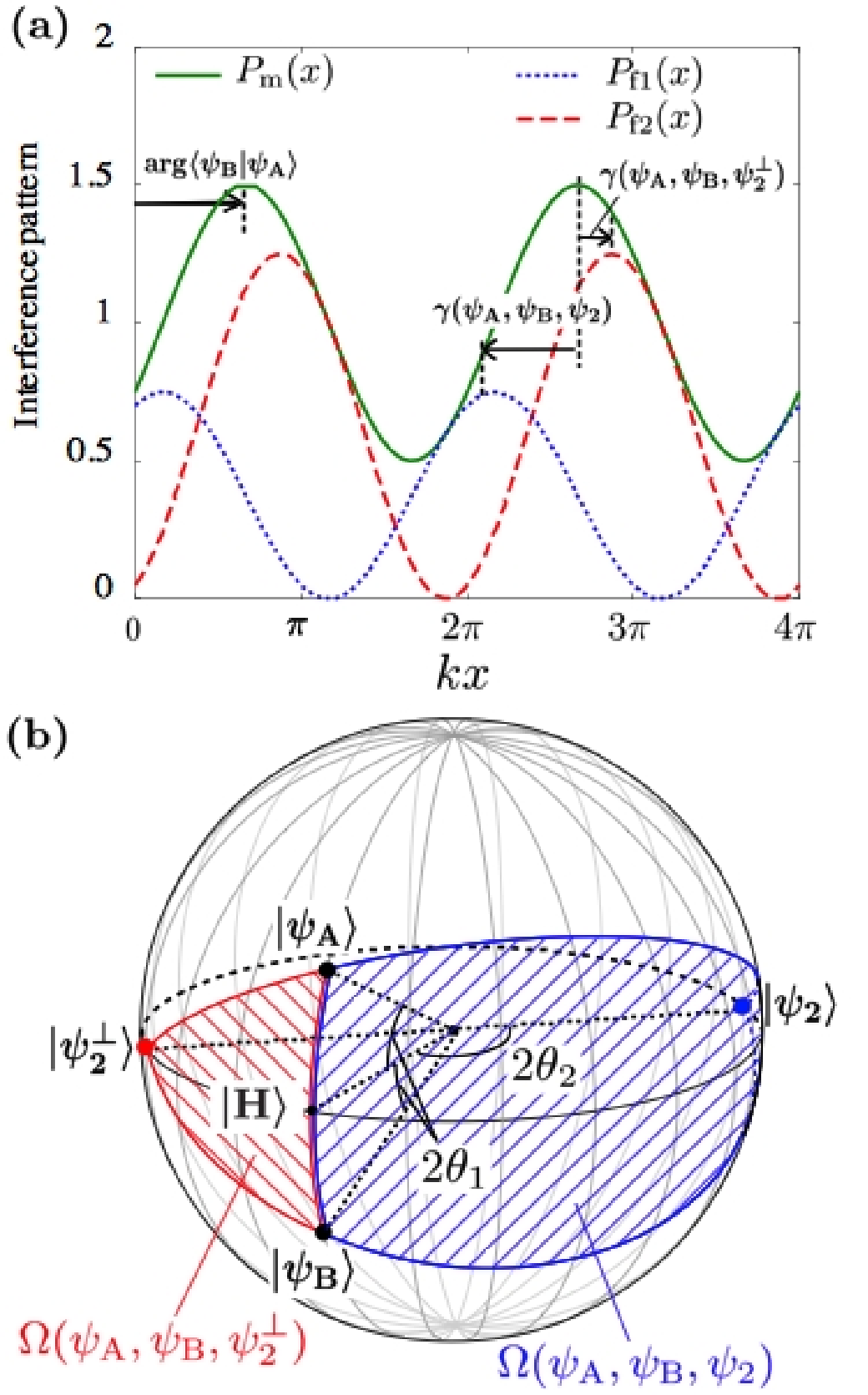}
\caption{(Color online). Separation of the partial interference pattern. 
(a) The partial interference pattern can be separated into two fringes
 with 100\% visibility shifted by $\gamma(\psi_\U{A},\psi_\U{B},\psi_2)$
 and $\gamma(\psi_\U{A},\psi_\U{B},\psi_2^\perp)$. (b) The shifts are
 related to the solid angles of the spherical triangles on the
 Poincar\'{e} sphere as
$\gamma(\psi_1,\psi_2,\psi_3)=-\frac{1}{2}\Omega(\psi_1,\psi_2,\psi_3)$.}
\label{fig:interference_split}
\end{center}
\end{figure}
As shown in the previous section, the which-path marker can destroy the
interference pattern effectively.
However, using the quantum eraser, the interference can be restored completely.
This implies that the complete interference pattern is buried under the
destroyed interference pattern.
In this section, we will show that the interference pattern destroyed by
the which-path marking contains two complete interference patterns
that are shifted by different amounts according to the projection of the
polarization state.
These phase shifts can be interpreted geometrically using the Poincar\'{e}
sphere, as shown below.

Equation~(\ref{eq:23}) can be separated into two terms using
the projected state $\ket{\psi_\U{2}}$ that satisfies
$|\bracketi{\psi_\U{2}}{\psi_\U{A}}|=|\bracketi{\psi_\U{2}}{\psi_\U{B}}|$
and its
orthogonal state $\ket{\psi_\U{2}^\perp}$ as follows:
\begin{align}
P_\U{m}(x)&=\bracketii{\Psi}{\left[
\left(\ket{\psi_\U{2}}\bra{\psi_\U{2}}+\ket{\psi_\U{2}^\perp}\bra{\psi_\U{2}^\perp}\right)
\otimes\ket{x}\bra{x}\right]}{\Psi}\nonumber\\
&\propto |c_\U{A}|^2P_\U{f1}(x)
+\left(1-|c_\U{A}|^2\right)P_\U{f2}(x),
\label{eq:1}
\end{align}
with
\begin{align}
P_\U{f1}(x)&\equiv
1+\cos\bigl[kx-\delta_\U{m}-\gamma(\psi_\U{A},\psi_\U{B},\psi_2)\bigr],\\
P_\U{f2}(x)&\equiv
1+\cos\bigl[kx-\delta_\U{m}-\gamma(\psi_\U{A},\psi_\U{B},\psi_2^\perp)\bigr],
\label{eq:12}
\end{align}
where the additional phase shift $\gamma$ is defined as
\begin{align}
\gamma(\psi_1,\psi_2,\psi_3)\equiv
\arg\bracketi{\psi_1}{\psi_2}\bracketi{\psi_2}{\psi_3}\bracketi{\psi_3}{\psi_1}.
\label{eq:2}
\end{align}
Due to the difference in the phase shifts,
$\gamma(\psi_\U{A},\psi_\U{B},\psi_2)-
\gamma(\psi_\U{A},\psi_\U{B},\psi_2^\perp)$,
even though both fringes $P_\U{f1}(x)$ and $P_\U{f2}(x)$ have
100\% visibility, the sum of these patterns has reduced visibility
[see Fig.~2(a)].

The right-hand side of Eq.~(\ref{eq:2}) is gauge-invariant, i.e.,
independent of the choice of the phase factor of each state, because the
bra and ket vectors for each state appear in a pair. This phase shift
$\gamma$ is identified with the Pancharatnam phase\cite{pancharatnam56:_proc}.
It can be shown that the Pancharatnam phase is proportional to 
the solid angle $\Omega(\psi_1,\psi_2,\psi_3)$ of the
spherical triangle connecting the states $\ket{\psi_1}$, $\ket{\psi_2}$,
and $\ket{\psi_3}$ with geodesic arcs on the Poincar\'{e} sphere
\cite{pancharatnam56:_proc,aravind92:_simpl_proof_of_panch_theor}, i.e., 
\begin{align}
\gamma(\psi_1,\psi_2,\psi_3)
=-\frac{1}{2}\Omega(\psi_1, \psi_2, \psi_3).
\label{eq:7}
\end{align}
Therefore, each phase shift of two fringes, $\gamma(\psi_\U{A},\psi_\U{B},\psi_\U{2})$ and
$\gamma(\psi_\U{A},\psi_\U{B},\psi_2^\perp)$, can be represented
geometrically on the Poincar\'{e} sphere. 
When the two marker states $\ket{\psi_\U{A}}$ and $\ket{\psi_\U{B}}$ are located
along a meridian symmetrically with respect to the equator,
the projected states $\ket{\psi_2}$ and $\ket{\psi_2^\perp}$ should be
on the equator in order to satisfy the condition 
$|\bracketi{\psi_2}{\psi_\U{A}}|=|\bracketi{\psi_2}{\psi_\U{B}}|$, as
shown in Fig.~2(b).

In particular, if $\ket{\psi_\U{A}}$ is perpendicular to $\ket{\psi_\U{B}}$, that is, 
a 100\%-effective which-path marker is
prepared, the four states $\ket{\psi_\U{A}}$,
$\ket{\psi_\U{B}}$, $\ket{\psi_2}$, and $\ket{\psi_2^\perp}$ lie
on the same great circle. Then, the following equation is satisfied:
\begin{align}
\gamma(\psi_\U{A},\psi_\U{B},\psi_2)
-\gamma(\psi_\U{A},\psi_\U{B},\psi_2^\perp)=\pi, 
\hspace*{0.1cm}|c_\U{A}|^2=\frac{1}{2}.
\end{align}
The total pattern is composed of two fringes having the same intensity
but opposite phases, and the interference is completely
washed out. 
A general case of partial erasure is shown in Fig.~\ref{fig:interference_split}.

\subsection{Nonlinear variation of Pancharatnam phase}
\begin{figure}[tbp]
\begin{center}
\includegraphics[width=8cm]{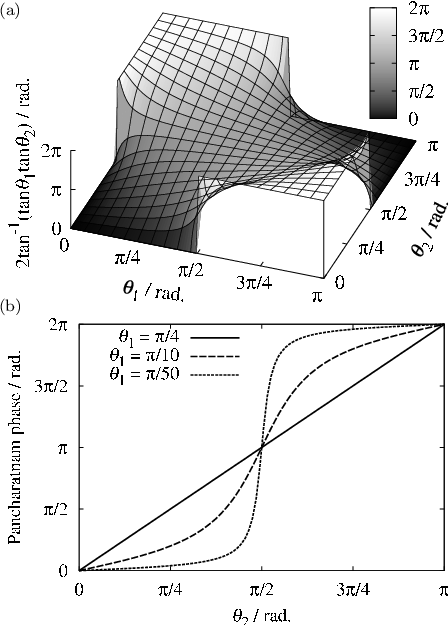}
\end{center}
\caption{Variation of the Pancharatnam phase (a) with respect to 
$\theta_1$ and $\theta_2$, and (b) with respect to $\theta_2$
for different $\theta_1$.} 
\label{fig:atan}
\end{figure}
From Eq.~(\ref{eq:7}), we can calculate the Pancharatnam phase in our
experiments.
We modify the standard double-slit interferometer to include two linear polarizers and
two quarter-wave plates, as shown in Fig.~\ref{fig:double-slit}(b). 

First, we prepare the initial polarization state $\ket{\psi_1}$
using the linear polarizer LP$_1:$
\begin{align}
\ket{\psi_1}=\cos\theta_1\ket{\U{H}}+\sin\theta_1\ket{\U{V}},
\end{align}
where $\theta_1$ is the angle between the horizontal line and the
transmission axis of LP$_1$; $\ket{\U{H}}$, the horizontal
polarization state; and $\ket{\U{V}}$, the vertical polarization
state. 
In our setup, the fast axes of two quarter-wave plates, QWP$_\U{A}$ and QWP$_\U{B}$,
are aligned to form angles of $0^\circ$ and $90^\circ$, respectively, from
the horizontal line.
Thus, they induce phase shifts of $\pm\pi/2$ between the horizontal and
the vertical components:
\begin{align}
\ket{\psi_\U{A}}&=\cos\theta_1\ket{\U{H}}+\ii\sin\theta_1\ket{\U{V}},\label{eq:4}\\
\ket{\psi_\U{B}}&=\ii\cos\theta_1\ket{\U{H}}+\sin\theta_1\ket{\U{V}}\label{eq:5}.
\end{align}
Here, the pair of quarter-wave plates serves as the which-path marker in Eq.~(\ref{eq:16}).
The final state of the polarization is expressed as $\ket{\psi_2}$:
\begin{align}
\ket{\psi_2}=\cos\theta_2\ket{\U{H}}+\sin\theta_2\ket{\U{V}},
\label{eq:6}
\end{align}
where $\theta_2$ is the angle between the horizontal line and the
transmission axis of LP$_2$. 

\begin{figure}[tbp]
\begin{center}
\includegraphics[width=6cm]{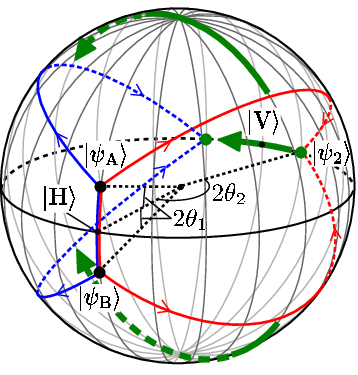}
\caption{(Color online). 
Geometrical interpretation of the nonlinear variation of the
 Pancharatnam phase around 
$(\theta_1,\theta_2)=(0,\pi/2)$. 
If $\ket{\psi_\U{A}}$ and $\ket{\psi_\U{B}}$ are
 close to each other on the Poincar\'{e} sphere, 
the area of the spherical triangle blows up very rapidly with the movement
 of $\ket{\psi_2}$ around $\theta_2=\pi/2$. 
}
\label{fig:poincare_sphere4}
\end{center}
\end{figure}
From Eqs.~(\ref{eq:4}), (\ref{eq:5}), and (\ref{eq:6}), we can obtain
the Pancharatnam phase as
\begin{align}
\gamma(&\psi_A,\psi_B,\psi_2)\nonumber\\
&=
\begin{cases}
2\tan^{-1}\bigl(\tan\theta_1\tan\theta_2\bigr)\hspace*{0.85cm}(\cos
 2\theta_1\geq 0),\\
2\tan^{-1}\bigl(\tan\theta_1\tan\theta_2\bigr)+\pi\hspace*{0.2cm}(\cos
 2\theta_1<0).
\end{cases}
\label{eq:3}
\end{align}
Figure \ref{fig:atan}(a) shows the variation of the first term of
Eq.~(\ref{eq:3}) with respect to $\theta_1$ and $\theta_2$.
It is noteworthy that this variation exhibits strong
nonlinearity around
$(\theta_1,\theta_2)=(0,\pi/2)$, $(\pi/2,0)$, $(\pi/2,\pi)$, and $(\pi,\pi/2)$. 
In Fig.~\ref{fig:atan}(b), the variation of the Pancharatnam phase 
with $\theta_2$ is plotted for different values 
of $\theta_1$. The figure shows that the smaller the value of $\theta_1$, 
the faster is the change in the Pancharatnam phase with
respect to $\theta_2$ near
$\theta_2=\pi/2$. 
We can observe this nonlinear variation as a rapid displacement of the
fringes when we change $\theta_2$ by rotating LP$_2$.

The nonlinear variation of the Pancharatnam phase can be explained by
the spherical geometry on the Poincar\'{e} sphere, as shown in
Fig.~\ref{fig:poincare_sphere4}. 
In our experiment, $\ket{\psi_\U{A}}$ and $\ket{\psi_\U{B}}$, given by
Eqs.~(\ref{eq:4}) and (\ref{eq:5}), respectively, can be depicted at a latitude of
$\pm 2\theta_1$ on the prime meridian, and the final state $\ket{\psi_2}$, given by 
Eq.~(\ref{eq:6}), can be depicted on the equator at a longitude of $2\theta_2$. 
We assume that $\ket{\psi_\U{A}}$ and $\ket{\psi_\U{B}}$ are located near
$\ket{\U{H}}$, that is, $0<\theta_1\ll\pi/4$ is satisfied, 
and $\ket{\psi_2}$ moves on the equator from $\ket{\U{H}}$. 
When the distance between $\ket{\psi_2}$ and $\ket{\U{V}}$ is greater
than $2\theta_1$, the area of the spherical triangle spanned by
$\ket{\psi_\U{A}}$, $\ket{\psi_\U{B}}$, and $\ket{\psi_2}$ remains
small. 
However, when $\ket{\psi_2}$
approaches $\ket{\U{V}}$ and the distance between them becomes lesser than
$2\theta_1$, the area of the spherical triangle increases very rapidly, and after traversing
$\ket{\U{V}}$, 
the triangle covers most of the Poincar\'{e} sphere.
This is the geometrical reasoning why the Pancharatnam phase
changes rapidly in certain conditions.

\section{Experiments}
\label{sec:experiments}
\begin{figure}[tbp]
\begin{center}
\includegraphics[width=8.5cm]{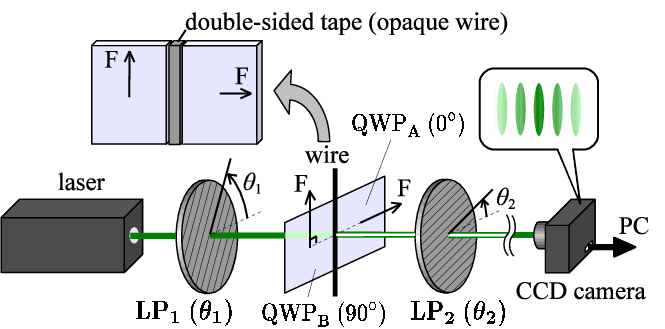}
\caption{(Color online). 
Experimental setup for double-slit quantum eraser. Light passing through the
 right and left of the wire interferes. Each path is marked by two
 film-type quarter-wave plates,
 QWP$_\U{A}$ and QWP$_\U{B}$, whose fast axes F make angles of 0$^\circ$ and
 90$^\circ$,
 respectively. The interference fringe is captured using a CCD camera.}
\label{fig:setup}
\end{center}
\end{figure}
The experimental setup is shown in Fig.~\ref{fig:setup}. 
The light source is a 532-$\U{nm}$ green laser with a 3-$\U{mm}$ beam
diameter (model DPGL-2200, SUWTECH). A thin opaque wire crossing the
beam works as the double slit; 
the light passing through the right- and left-hand sides of the wire
interferes due to diffraction.
We attached two film-type quarter-wave plates having the orthogonal
fast axes, $0^\circ$ and $90^\circ$, with a thin piece of double-sided
adhesive tape that works as a wire. 
A double slit having a similar design has been introduced by Hilmer and Kwiat
\cite{hillmer07:_do_it_yousel_quant_eraser}.

Two film-type linear polarizers, LP$_1$ and LP$_2$, are attached to the rotatable
mounts with graduated scales for adjusting the angles $\theta_1$ and $\theta_2$.
At a distance of approximately 1\,m from the double slit, the recombined beam
is captured using a charge-coupled device (CCD) camera (model LBP-2-USB, Newport) 
connected to a personal computer (PC). The CCD camera has a resolution
of 640$\times$480 pixels, each having a size of $9$\,$\U{um}\times
8$\,$\U{um}$, and it is equipped with a gain controller. 

\subsection{Experimental results of quantum erasers}
\begin{figure}[tbp]
\begin{center}
\includegraphics[width=6cm]{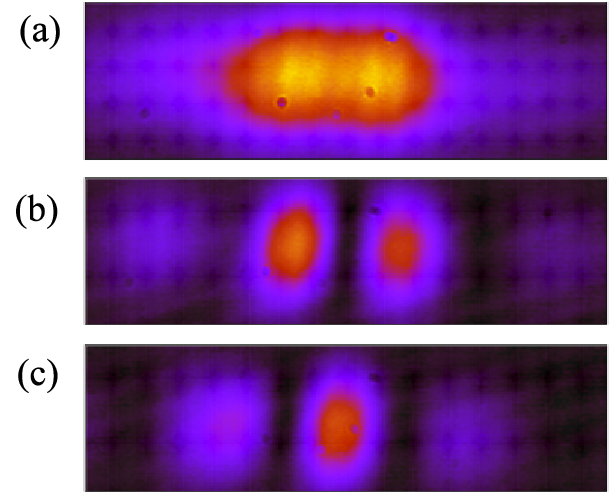}\\
\end{center}
\caption{(Color online). 
Interference patterns captured using a CCD camera. (a) By setting
 $\theta_1=\pi/4$ and removing LP$_2$, a typical diffraction
 pattern is observed. (b) By setting
 $\theta_1=\pi/4$ and $\theta_2=0$, the fringe pattern reappears. 
(c) By setting $\theta_1=\pi/4$ and $\theta_2=\pi/2$, the fringe pattern is
 out of phase with the case of $\theta_2=0$.}
\label{fig:fringe_in_eraser}
\end{figure}
\begin{figure}[tbp]
\begin{center}
\includegraphics[width=8cm]{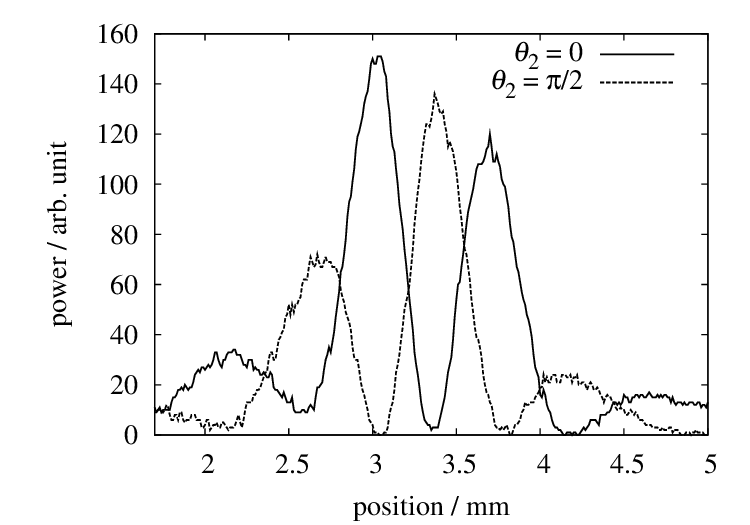}
\caption{Recovered interference fringes for the quantum eraser with $\theta_2=0$ and $\theta_2=\pi/2$}
\label{fig:45-90_45-0}
\end{center}
\end{figure}

First, by setting $\theta_1=\pi/4$ and removing the linear polarizer LP$_2$, the initial state
of polarization $\ket{\U{D}}$ is evolved into two orthogonal states
through the quarter-wave plates, right circular
polarization, and left circular polarization according to the
paths. 
Because we can determine which slit the photon has passed through by measuring
the polarity of the circular polarization of the photon, no interference pattern
is obtained. 
(This is mathematically confirmed from Eq.~(\ref{eq:9}), which
vanishes when $\ket{\psi_\U{A}}$ is orthogonal to $\ket{\psi_\U{B}}$.)
We observed a typical diffraction pattern that only has
broad peaks, as shown in Fig.~\ref{fig:fringe_in_eraser}(a). 

By inserting LP$_2$, the right and left circular polarizations are projected into the same linear
polarization with the same probability, and therefore, the polarization
provides no which-path information. As a result, the interference fringe reappears.
(Mathematically, this corresponds to the fact that Eq.~(\ref{eq:25}) becomes
unity when $|c_\U{A}|=|c_\U{B}|$.) Figure~\ref{fig:fringe_in_eraser}(b) shows the
recovered interference fringe for $\theta_2=0$. 
Similarly, for 
$\theta_2=\pi/2$, we can obtain the corresponding interference fringe, as shown in Fig.~\ref{fig:fringe_in_eraser}(c), 
which is out of phase with that observed for $\theta_2=0$ (see
Fig.~\ref{fig:45-90_45-0}). 
This phase difference is attributed to the Pancharatnam phase. 
The sum of these interference patterns reproduces the broad
peak pattern, as shown in Fig.~\ref{fig:fringe_in_eraser}(a), that is
obtained in the absence of  LP$_2$.
Therefore, the quantum eraser actually filters out one of these
fringes and perfectly recovers the visibility.

\subsection{Observation of Pancharatnam phase and its nonlinearity}
\begin{figure}[tbp]
\begin{center}
\includegraphics[width=8.5cm]{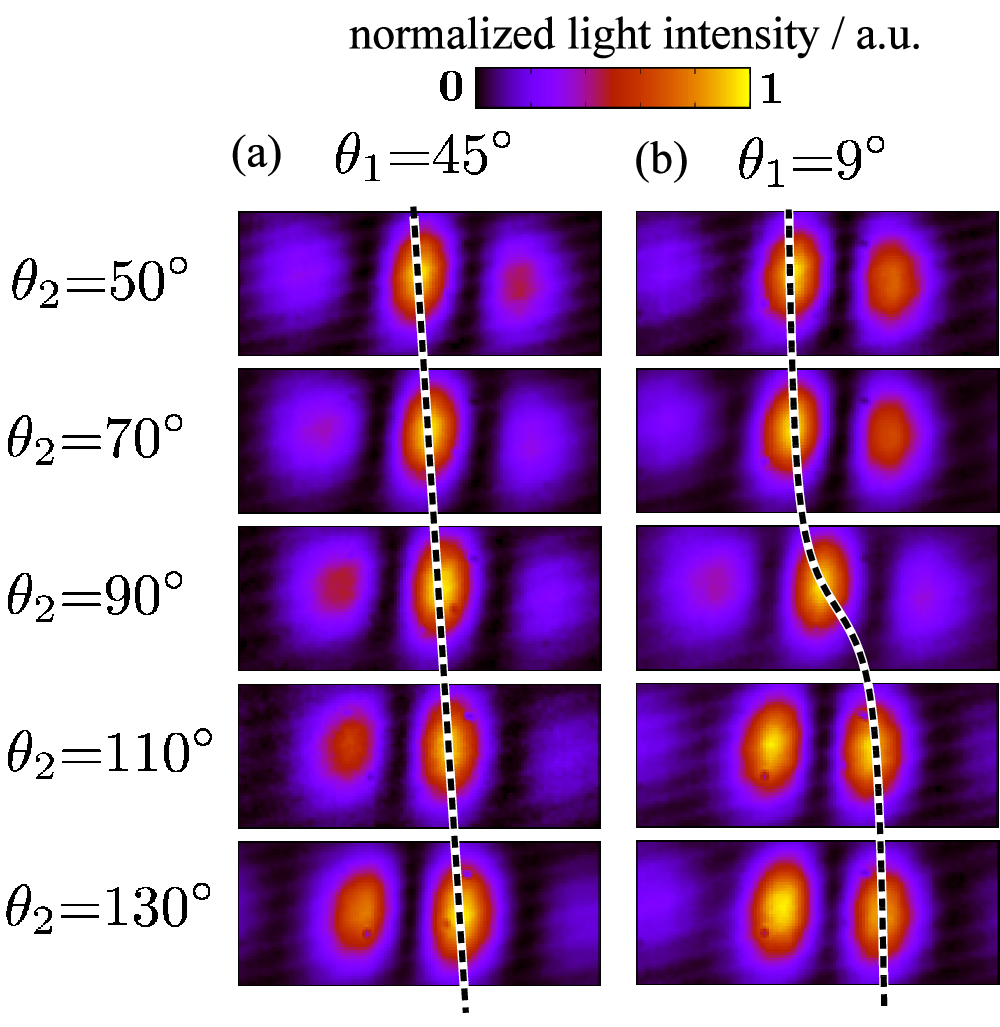}
\caption{(Color online). 
The shift of fringes induced by the Pancharatnam phase with respect
 to $\theta_2$ (a) when
 $\theta_1=45^\circ$ and (b) when $\theta_1=9^\circ$. The light
 intensity of each frame is normalized individually. When
 $\theta_1=9^\circ$, the fringe exhibits a rapid displacement around $\theta_2=90^\circ$.}
\label{fig:Pancharatnam_shift}
\end{center}
\end{figure}
\begin{figure}[tbp]
\begin{center}
\includegraphics[width=8.5cm]{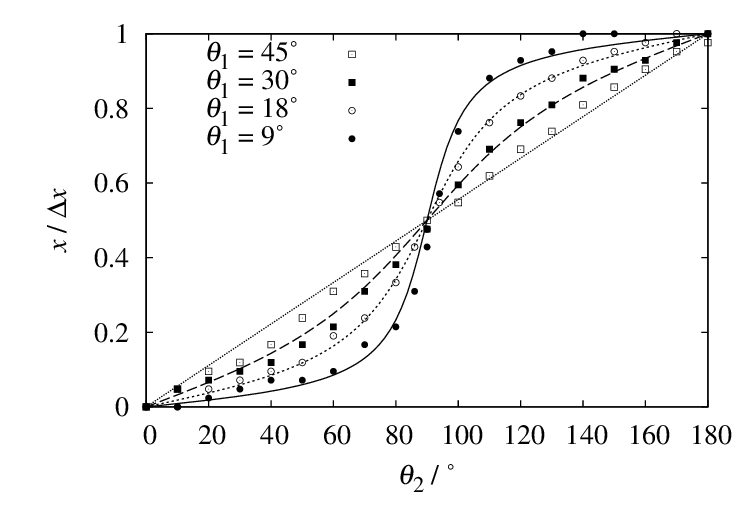}
\end{center}
\vspace*{-0.7cm}
\caption{Experimental results of Pancharatnam phase with respect to $\theta_2$ for
 different $\theta_1$. The vertical axis shows the displacement of the 
fringe normalized by the spatial period of the fringe.}
\label{fig:ex_pancharatnam}
\end{figure}
In order to observe the variation of the Pancharatnam phase with respect
to $\theta_1$ and $\theta_2$, we measured
the displacement of the fringes. 
Figure~\ref{fig:Pancharatnam_shift} shows the fringe shift with respect
to $\theta_2$ for fixed $\theta_1$.
The light intensity of each fringe in Fig.~\ref{fig:Pancharatnam_shift}
is normalized individually.
When $\theta_1=45^\circ$, the fringe moves linearly with
respect to the change in $\theta_2$, 
as shown in Fig.~\ref{fig:Pancharatnam_shift}(a). However, setting
$\theta_1=9^\circ$, the
fringe exhibits a quick displacement around $\theta_2=90^\circ$, as shown in Fig.~\ref{fig:Pancharatnam_shift}(b).

Figure~\ref{fig:ex_pancharatnam} shows the variation of the Pancharatnam
phase 
with respect to $\theta_2$ for $\theta_1=45^\circ$,
$30^\circ$, $18^\circ$, and $9^\circ$.
The points in Fig.~\ref{fig:ex_pancharatnam} indicate the
experimental results and the solid lines indicate the theoretical lines
calculated using Eq.~(\ref{eq:3}). 
The vertical axis is the
displacement of the fringe $x$ normalized by the spatial period of the
fringe $\Delta x$. 
The origin of the vertical axis is determined by the position of 
the fringes when 
$\theta_2=0^\circ$.
All the experimental results agree well with
the theoretical ones. The gradient of the variation of the shift around
$\theta_2=90^\circ$ increases as $\theta_1$
is decreased. This implies that the variation of the shift becomes more
sensitive to the variation of 
the last polarization state.

\section{Summary}
\label{sec:summary-discussion}
We have shown that the Pancharatnam phase manifests in the
setup for quantum erasers. 
In our experiment, we have introduced 
a double-slit interferometer with internal states of
a photon and demonstrated which-path marking, quantum erasers, and the
variation of the geometric phases. 
The visibility of the interference fringe is related to the which-path
marking and the quantum eraser, and the phase shift of the
interference shows the manifestation of the Pancharatnam phase.
Moreover, we have demonstrated that the Pancharatnam phase could become sensitive
to a change in the polarization state. 
This fact can be utilized 
for high-precision measurement of the polarization.

Even though our experiment is performed with classical light, it
serves the purpose of showing the quantum-mechanical meaning of which-path marking, quantum
erasers, and geometric phases since photons are noninteracting Bose
particles and our tests can
be straightforwardly extended to experiments with single photons\cite{chiao86:_manif_of_berry_topol_phases_for_photon}. 

\begin{acknowledgments}
We gratefully thank Youhei Hosoe for assisting us in our experiments.
We thank Yosuke Nakata and Rikizo Ikuta for providing useful comments
 and suggestions.
This research is supported by the global COE program ``Photonics and
 Electronics Science and Engineering'' at Kyoto University.
\end{acknowledgments}

\end{document}